\documentclass[8.5pt,twoside,twocolumn]{article}
\usepackage[super,sort&compress,comma]{natbib}
\usepackage{color}
\usepackage[version=3]{mhchem} 
\usepackage{times,mathptmx}
\usepackage{xspace}
\usepackage{sectsty}
\usepackage{balance} 

\usepackage{graphicx} 
\usepackage{lastpage}
\usepackage[format=plain,justification=raggedright,singlelinecheck=false,font=small,labelfont=bf,labelsep=space]{caption} 
\usepackage{fancyhdr}
\usepackage[scanall]{psfrag}
\usepackage{bm}

\begin{document}

\thispagestyle{plain}
\fancypagestyle{plain}{
\fancyhead[L]{\includegraphics[height=8pt]{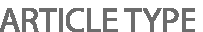}}
\fancyhead[C]{\hspace{-1cm}\includegraphics[height=20pt]{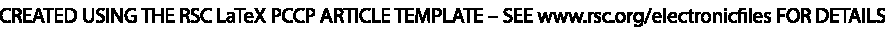}}
\fancyhead[R]{\includegraphics[height=10pt]{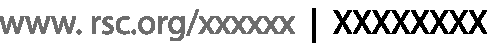}\vspace{-0.2cm}}
\renewcommand{\headrulewidth}{1pt}}
\renewcommand{\thefootnote}{\fnsymbol{footnote}}
\renewcommand\footnoterule{\vspace*{1pt}%
\hrule width 3.4in height 0.4pt \vspace*{5pt}} 
\setcounter{secnumdepth}{5}

\makeatletter 
\def\subsubsection{\@startsection{subsubsection}{3}{10pt}{-1.25ex plus -1ex minus -.1ex}{0ex plus 0ex}{\normalsize\bf}} 
\def\paragraph{\@startsection{paragraph}{4}{10pt}{-1.25ex plus -1ex minus -.1ex}{0ex plus 0ex}{\normalsize\textit}} 
\renewcommand\@biblabel[1]{#1}            
\renewcommand\@makefntext[1]%
{\noindent\makebox[0pt][r]{\@thefnmark\,}#1}
\makeatother 
\renewcommand{\figurename}{\small{Fig.}~}
\sectionfont{\large}
\subsectionfont{\normalsize} 

\fancyfoot{}
\fancyfoot[LO,RE]{\vspace{-7pt}\includegraphics[height=9pt]{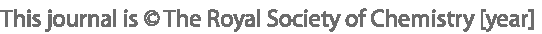}}
\fancyfoot[CO]{\vspace{-7.2pt}\hspace{12.2cm}\includegraphics{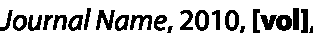}}
\fancyfoot[CE]{\vspace{-7.5pt}\hspace{-13.5cm}\includegraphics{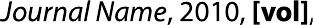}}
\fancyfoot[RO]{\footnotesize{\sffamily{1--\pageref{LastPage} ~\textbar  \hspace{2pt}\thepage}}}
\fancyfoot[LE]{\footnotesize{\sffamily{\thepage~\textbar\hspace{3.45cm} 1--\pageref{LastPage}}}}
\fancyhead{}
\renewcommand{\headrulewidth}{1pt} 
\renewcommand{\footrulewidth}{1pt}
\setlength{\arrayrulewidth}{1pt}
\setlength{\columnsep}{6.5mm}
\setlength\bibsep{1pt}

\newcommand\jacs{J. Am. Chem. Soc.}
\newcommand\jcp{J. Chem. Phys.}
\newcommand\prsa{Proc. R. Soc. A}
\newcommand\anrb{Ann. Rev. Biophys.}
\newcommand\pnasu{Proc. Natl. Acad. Sci. USA}
\newcommand\prl{Phys. Rev. Lett.}
\newcommand\pre{Phys. Rev. E}
\newcommand\pra{Phys. Rev. A}
\newcommand\prb{Phys. Rev. B}
\newcommand\ic{Inorg. Chem.}
\newcommand\jpcb{J. Phys. Chem. B}
\newcommand\jpcm{J. Phys. Cond. Mat.}
\newcommand\acp{Adv. Chem. Phys.}
\newcommand\jpca{J. Phys. Chem. A}
\newcommand\jpcc{J. Phys. Chem. C}
\newcommand\jpc{J. Phys. Chem.}
\newcommand\pccp{Phys. Chem. Chem. Phys.}
\newcommand\ptrsla{Philos. Trans. Roy. Soc. Lond. A}
\newcommand\jcsft{J. Chem. Soc., Faraday Trans.}
\newcommand\jcscc{J. Chem. Soc., Chem. Commun.}

\twocolumn[
  \begin{@twocolumnfalse}
\noindent\LARGE{\textbf{Energy Landscapes, Structural Topologies and Rearrangement \\
Mechanisms in Clusters of Dipolar Particles}}
\vspace{0.6cm}

\noindent\large{\textbf{James D.~Farrell,$^a$ Christabel Lines,$^a$ James J.~Shepherd,$^a$
Dwaipayan Chakrabarti,$^a$ Mark A.~Miller$^{ab}$ and David J.~Wales$^a$}}\vspace{0.5cm}

\noindent\textit{\small{\textbf{Received Xth XXXXXXXXXX 20XX, Accepted Xth XXXXXXXXX 20XX\newline
First published on the web Xth XXXXXXXXXX 200X}}}

\noindent \textbf{\small{DOI: 10.1039/b000000x}}
\vspace{0.6cm}
\noindent \normalsize{

Clusters of spherical particles with isotropic attraction favour compact structures that
maximise the number of energetically optimal nearest-neighbour interactions.  In contrast, dipolar
interactions lead to the formation of chains with a low coordination number.  When both
isotropic and dipolar interactions are present, the competition between them can lead to
intricate knot, link and coil structures.  Here, we investigate how these structures may
self-organise and interconvert in clusters bound by the Stockmayer potential (Lennard-Jones
plus point dipole). We map out the low-lying region of the energy landscape 
using disconnectivity graphs to follow how it evolves as the strength of the dipolar interactions 
increases.   From comprehensive
surveys of isomerisation pathways, we identify a number of rearrangement mechanisms that
allow the topology of chain-like structures to interconvert.
}
\vspace{0.5cm}
 \end{@twocolumnfalse}
  ]
\footnotetext{$^a$University Chemical Laboratory, Lensfield Road, Cambridge CB2 1EW, United Kingdom.
Fax: 01223 336362; \
Tel: 01223 336300; \
E-mail: jdf43@cam.ac.uk, mam1000@cam.ac.uk, dw34@cam.ac.uk}
\footnotetext{$^b$Current address: Department of Chemistry, Durham University, South Road, Durham DH1 3LE,
United Kingdom.  E-mail: m.a.miller@durham.ac.uk}

\section{Introduction}

The field of chemical topology\cite{FrischW61} can be traced back to the first characterisation
of interlocked
organic rings and the introduction of the term catenane by Wasserman in 1960.\cite{Wasserman60}
The two rings in a catenane are not chemically bonded, and the topology must therefore be
specified to distinguish the structure from the unlinked rings, giving rise to the concept
of topological isomers.  Wasserman's short report also points out that a single cyclic
hydrocarbon of sufficient length may, in principle, exist not only in the form of a
simple loop but instead as a topologically distinct knot.

Since this early work,
there has been considerable progress both in the theoretical
understanding\cite{MansfieldD10a,MansfieldD10b} of ``mechanically linked''
cyclic molecules and in efficient methods for synthesising them.\cite{barinFS12}
Knotted topologies have even
been discovered in organic synthesis even when originally unexpected.\cite{Ponnuswamy12a}
The prevalence and importance of knots in biological macromolecules has also been
recognised.  Catenated and knotted forms of circular DNA naturally occur, for which 
topological isomers are enzymatically interconverted during
replication, transcription, and recombination.\cite{WassermanC86,MeluzziSA10}
Knot-like motifs arise in the native structure of a surprisingly large number of proteins.
Although proteins are unbranched chains---and therefore have a topology that is
formally trivial---considerable insight into the role of knots can be gained by
chemically joining the ends of the protein to make a cyclic topology that contains
a genuine topological knot.\cite{Mallam10a}

Cyclic structures can also emerge from the self-assembly of particles that interact
{\it via} non-covalent interactions.  In particular, colloidal particles that
carry a dipole moment tend to form chains that can close into loops,\cite{Klokkenburg06a}
and the possibility of knotted topologies in such systems has been pointed out.\cite{Keng07a}
Highly intricate knotted topologies also arise in other types of colloids
in a way that is conceptually different---not from chains
of particles themselves, but from lines of defects in cholesteric
liquid crystals, where twisted
nematic order is disrupted by anchoring to spherical particles suspended in the
solution.\cite{JampaniSRCZM11,TkalecRSZM11}  Once again, however, it is the
anisotropic interactions between particles that leads to
the existence of these lines and the relevance of topological considerations.

Global optimisation calculations have shown that knots, links and coils are energetically
favourable structures for clusters of Stockmayer particles,
which possess a permanent dipole plus an isotropic soft core and
attractive tail.\cite{MillerW05}  These intricate topologies arise from the competition
between the dipole--dipole interactions, which promote chain formation, and the isotropic
attraction, which favours compact, highly-coordinated structures.  The energetically
optimal compromise is often a continuous closed-loop chain, which is entwined
with itself or with another loop in order to increase the number of nearest-neighbour contacts.

Most of the literature concerning the Stockmayer model has addressed the structure and
thermodynamics of the bulk fluid phase.  In this context, too, the interplay between chains and
compact structures is crucial.  Van Leeuwen and Smit were amongst the first to point out that
gas--liquid phase separation is suppressed if the dipolar attraction
dominates.\cite{vanLeeuwen93a}  Ten Wolde {\it et al.}~showed that droplets of Stockmayer
particles appear from the vapour phase first by formation of chains and then by collapse of
the chains into globules that nevertheless retain a chain-like structure.\cite{tenWolde1999}
The more general field of dipolar fluids, including the Stockmayer model, has been
summarised in an informative review by Teixeira {\it et al.}\cite{Teixeira00a}
There has been relatively little work on finite clusters bound by the Stockmayer
potential,\cite{Lavender94a} though the model has now been recognised as having applications
in self-assembly.\cite{vanWorkum06a}

The potential for a cluster of $N$ Stockmayer particles takes the form
\begin{equation}
\label{eq:stockmayer}
\begin{split}
V=\epsilon \sum_{i<j}^{N}&
\left\{
	4
	\left[
		\left(
			\frac{\sigma}{r_{ij}}
		\right)^{12} -
		\left(
			\frac{\sigma}{r_{ij}}
		\right)^{6}
	\right] \right. \\
&	\left. +\frac{\mu^{2}\sigma^{3}}{r_{ij}^{3}}
	\left[
		\bm{\hat{\mu}}_{i} \cdot \bm{\hat{\mu}}_{j} - \frac{3}{r_{ij}^{2}}
		\big(
			\bm{\hat{\mu}}_{i} \cdot \bm{r_{ij}}
		\big)
		\big(
			\bm{\hat{\mu}}_{j} \cdot \bm{r_{ij}}
		\big)
	\right]
\right\},
\end{split}
\end{equation}
where $\bm{r_{ij}}$ is the position vector of particle $i$ with respect to particle $j$,
$\bm{\hat{\mu}}_{i}$ is a unit vector along the dipole moment of particle $i$, $\epsilon$
and $\sigma$ are the Lennard-Jones (LJ) units of energy and distance, respectively, and
$\mu$ is a dimensionless parameter that determines the relative strength of the dipolar and
LJ components of the potential.  As $\mu$ increases and the dipolar contribution to the
energy dominates, the Lennard-Jones well depth $\epsilon$ is no longer a convenient unit of
energy.  Here, we will report energies in units of the well-depth $\epsilon^*$ of the full
Stockmayer pair potential for parallel head-to-tail dipole vectors at the relevant value of
$\mu$.

The structure of the clusters is tuned by the dipole moment strength $\mu$.
In the limit of a weak dipole (small $\mu$), compact, icosahedral structures are
favoured.\cite{Hoare79a}  For strong dipoles the particles behave as dipolar soft spheres,
producing chains, rings, and branched clusters, such as those formed
by assemblies of ferromagnetic nanoparticles.\cite{DingLZZWSW12}
It is in the intermediate regime, where competition between the isotropic and directional 
contributions leads to frustration, and interesting topologies with as many as ten
irreducible crossings may arise.\cite{MillerW05}

The current work builds on these findings concerning energetically optimal structures by making
a broader survey of the energy landscape for a selection of cases.  The analysis presented here
enables us to characterise the overall evolution of the energy landscape between the 
Lennard-Jones and dipole-dominated limits, and to identify pathways and rearrangement mechanisms 
by which different morphologies interconvert.  Some of the key properties can be observed in the 
small St$_{13}$ cluster (where St$_{N}$ denotes an assembly of $N$ Stockmayer particles).
The analysis is then extended to more complex rearrangements of knotted morphologies in
the St$_{21}$ and St$_{38}$ clusters.

\section{Methods}

\subsection{Optimisation}
A Stockmayer particle consists of a single interaction site with an
isotropic Lennard-Jones component and an anisotropic point dipole.
Since the dipole has cylindrical symmetry, each particle has five degrees
of freedom: three translational and two rotational.  In previous
work\cite{MillerW05} on the global optimisation of Stockmayer particles,
the translational coordinates were represented in Cartesians and the 
orientation of the dipole was described using the polar angle $\theta$ and 
azimuthal angle $\phi$.

The present work requires optimisation of first-order saddle points and
the calculation of pathways, as well as characterisation of minima on the
potential energy surface (PES).  For such applications it is more 
convenient to represent the orientation of the particles using the general 
angle-axis framework,\cite{Chakrabarti09a,Wales05a} where the orientation
of a rigid body is obtained by rotating a particle from a reference 
orientation about an axis specified by the rotation vector 
${\bf p}$ through an angle given by its magnitude $|{\bf p}|$.  Although 
this representation introduces a redundant sixth degree of freedom for each 
particle in this case, it has the advantage that the components of ${\bf p}$ 
can be treated as unbounded variables, unlike $\theta$ and $\phi$.  
The redundant degrees of freedom contribute additional zero eigenvalues to
the Hessian matrix; the corresponding eigenvectors can be obtained 
analytically\cite{Chakrabarti09a} and are used for projection while 
characterising the transition state as described later.  

In the case of Stockmayer particles, we arbitrarily took the reference orientation
of the dipole along the laboratory-fixed $z$ axis.  The unit
dipole vector ${\bm{\hat\mu}}_i$ of a particle can then be obtained from the
angle-axis variables ${\bf p}_i$.  All angular first and second derivatives 
of $V$ can then be written with respect to the components of ${\bf p}_i$.

\subsection{Calculating pathways\label{pathways}}

A potential energy landscape can be characterised by mapping out its local minima,
which correspond to locally stable inherent structures,\cite{Stillinger83a}
and their connections via first-order saddle points (transition states).

The survey of an energy landscape typically starts from a small selection of
energetically low-lying minima obtained from a basin-hopping\cite{Wales97a,GMIN}
global optimisation calculation.  To find the pathways that connect pairs of
these structures,\cite{CarrTW05} we employ the doubly-nudged elastic band 
method,\cite{TrygubenkoW04} as implemented in the {\tt OPTIM} package.\cite{OPTIM} 
In this approach, a discretised interpolation is first established between the two 
minima in the full configuration space. The path is then allowed to relax on the 
PES while the states are kept approximately evenly spaced along the path by springs.
Local maxima on the path are then tightly converged to transition states using
hybrid eigenvector-following,\cite{HenkelmanJ99,MunroW99,KumedaWM01} and
the connectivity is determined by approximate steepest-descent minimisations, adding 
any new minima thus found to the database of structures. It is common for new minima 
to be found during this procedure and so the first iteration of attempted connections 
rarely produces a fully connected path between the minima that were originally specified.

Missing sections on a path are approached in the same way.  Candidate pairs of
minima to be bridged are selected by repeated application of the missing connection
algorithm,\cite{CarrTW05} also implemented in the {\tt OPTIM} package,\cite{OPTIM} 
until a connected sequence of minima and transition states (a discrete path) is found.
To characterise complete paths, the minima are considered to be nodes on a graph and the weight 
of the edge between two minima is set to zero if they are already directly connected by 
a single transition state in the database. If no connection is known, the weight is set 
to a function of the shortest Euclidean distance between the minima.\cite{CarrTW05}
Hence, the algorithm identifies breaks in the path that are short in configuration
space, making it more plausible that a connection will be found.  The Euclidean
distance between different Stockmayer clusters was determined from the positions of the 
particles only, ignoring the orientations of the dipole moments.

\subsection{Discrete Path Sampling}

Once an initial path has been established, the stationary point database must be
expanded in order to obtain a more complete picture of the landscape.
For small systems, a comprehensive survey of the landscape is possible by
repeatedly attempting to connect all pairs of minima already in the database by the
method described in Section \ref{pathways} until no new minima or pathways
emerge.  This approach was employed for the St$_{13}$ clusters.

For clusters even just a few particles larger,
it is not feasible to obtain exhaustive lists of stationary points.
It is then most informative to obtain a good representation of the regions surrounding the
lowest-lying minima of each structural family and a set of kinetically relevant pathways
between them.  Algorithms for connecting regions of configuration space in this way 
using discrete path sampling\cite{Wales02,Wales04,TrygubenkoW06a} (DPS) have been described
in detail in previous work.\cite{CarrW05,StrodelWW07}
Consider two regions on the landscape, A and B, which contain the minima we wish to connect.
Minima in neither region that nevertheless appear on pathways between A and B minima belong
to the intervening region, denoted I.  Assuming Markovian dynamics, that local equilibrium is
reached within the A and B sets, and that the steady-state approximation applies to minima
in the I set, the steady-state rate constants $k_{\rm BA}^{\rm SS}$ and $k_{\rm AB}^{\rm SS}$
can each be written as a sum over discrete paths.\cite{Wales02}
In turn, the contribution of a given discrete path can be expressed in terms of the
individual rate constants $k_{ij}$ for each successive pair $(i,j)$ of minima along
the path.

Since we are concerned with characterising the PES rather than obtaining quantitative
values for the rate constants, we estimate the single-step rate constants $k_{ij}$
using a simplified version of harmonic transition state theory, where the ratio of normal
mode frequencies at the transition state and the minimum is set to unity.  This simplification 
has the advantage of making it unnecessary to consider the relationship between the vibrational
and librational contributions to the normal modes of the Stockmayer particles.  To proceed 
otherwise would require specialisation of the model by choosing a particular moment of inertia for
the particles.  With this simplification, the expression for the rate constant reduces to
\begin{displaymath}
k_{ij}=\frac{o_i}{o^\dag_{ij}}\exp[-(V^\dag_{ij}-V_i)/(k_{\rm B}T)],
\end{displaymath}
where $o_{i}$ denotes the order of the point group for minimum $i$, and the superscript $\dag$ 
is used to refer to the transition state.  For the calculation of rate constants, we work at a
temperature of $k_{\rm B}T/\epsilon^*=1/30$, which is the regime probed by some recent
experiments in which rings of magnetic dipolar colloids have been
observed.\cite{DingLZZWSW12}

Using the simplified rate constants for individual steps, the discrete path
that makes the largest contribution to $k_{BA}^{SS}$ (the fastest path)
can be identified.
Attempts are then made to shortcut this path by directly connecting minima that lie on it
but are separated by intervening minima.  The choice of minima can be made on the basis of their 
separation along the path or in Euclidean space, or from the height of the barrier that separates them.
These schemes guide the exploration of the PES towards
kinetically relevant minima and transition states,
and are here applied to the more complex landscapes of the St$_{21}$ and St$_{38}$ clusters.

It is possible for artificial kinetic traps to arise as the database is generated---that is,
for minima to be connected to the A and B regions by large barriers, where smaller barriers exist
but have not yet been found, resulting in spuriously small rate constants.
As the database grows, we periodically attempt to remove such traps
by identifying pairs of minima based on the ratio of the potential energy barrier (from minimum to
transition state) to the potential energy difference between the two minima connected
by the transition state.\cite{StrodelWW07}  These pairs are then subjected to reconnection
searches as described above. Such `untrapping' cycles were performed for all databases until the 
low-energy region of the landscape converged.

The methods described above are implemented in the program {\tt PATHSAMPLE},\cite{PATHSAMPLE}
which is a driver for {\tt OPTIM},\cite{OPTIM}
as well as a tool for analysis of the resulting kinetic transition networks.

\subsection{Topological Characterisation}

To determine the topology of a given structure, it is first necessary to trace chains of
head-to-tail dipoles within the cluster.  An intuitive and robust method for
identifying connectivity has been described by Miller and Wales.\cite{MillerW05}
Beginning with a particle $i$, the next particle in the chain is the one with the lowest (most favourable)
dipole--dipole interaction energy with $i$, located in the half-space into which the dipole at $i$ points.
Similarly, the previous particle is the one with the lowest interaction energy in the complementary half-space.
Connections in the chains that this procedure identifies will be referred to here as `bonds.'
\par
In some structures, the bonds define a closed-loop chain that incorporates all the particles.
If this loop cannot be unravelled into a trivial circle containing no crossings without breaking
any bonds, then the structure is a knot.  A knot is classified in the Rolfsen notation\cite{Rolfsen76}
according to the two-dimensional projection of the chain that contains the smallest possible
number of crossings.  For the structures encountered in this work, the topology can be determined
from an arbitrary projection by considering all combinations of splittings at the crossings and
thereby evaluating the knot's Jones polynomial.\cite{Jones85a}  The Rolfsen symbol can then be
determined from tables.\cite{Adams04a}

\section{Results and Discussion}
\subsection{St$_{13}$ clusters}
For St$_{13}$, four structures successively become the global minimum as $\mu$ is increased
from zero: a distorted icosahedron, with $D_{3}$ symmetry; a
hexagonal antiprism, with $D_{6d}$ symmetry; stacked rings of six and seven particles, with $C_{s}$ symmetry; and a planar
thirteen-particle ring, with $D_{13h}$ symmetry.  The point groups here refer to symmetry operations that map the particle
positions onto each other without considering the direction of the dipoles.

Databases of minima and transition states have been constructed for the three values of $\mu$ at which the global minimum changes,
1.42, 2.54, and 2.66, to show competition between the different morphologies,
\begin{figure*}[!tb]
\centerline{\includegraphics[width=0.99\textwidth]{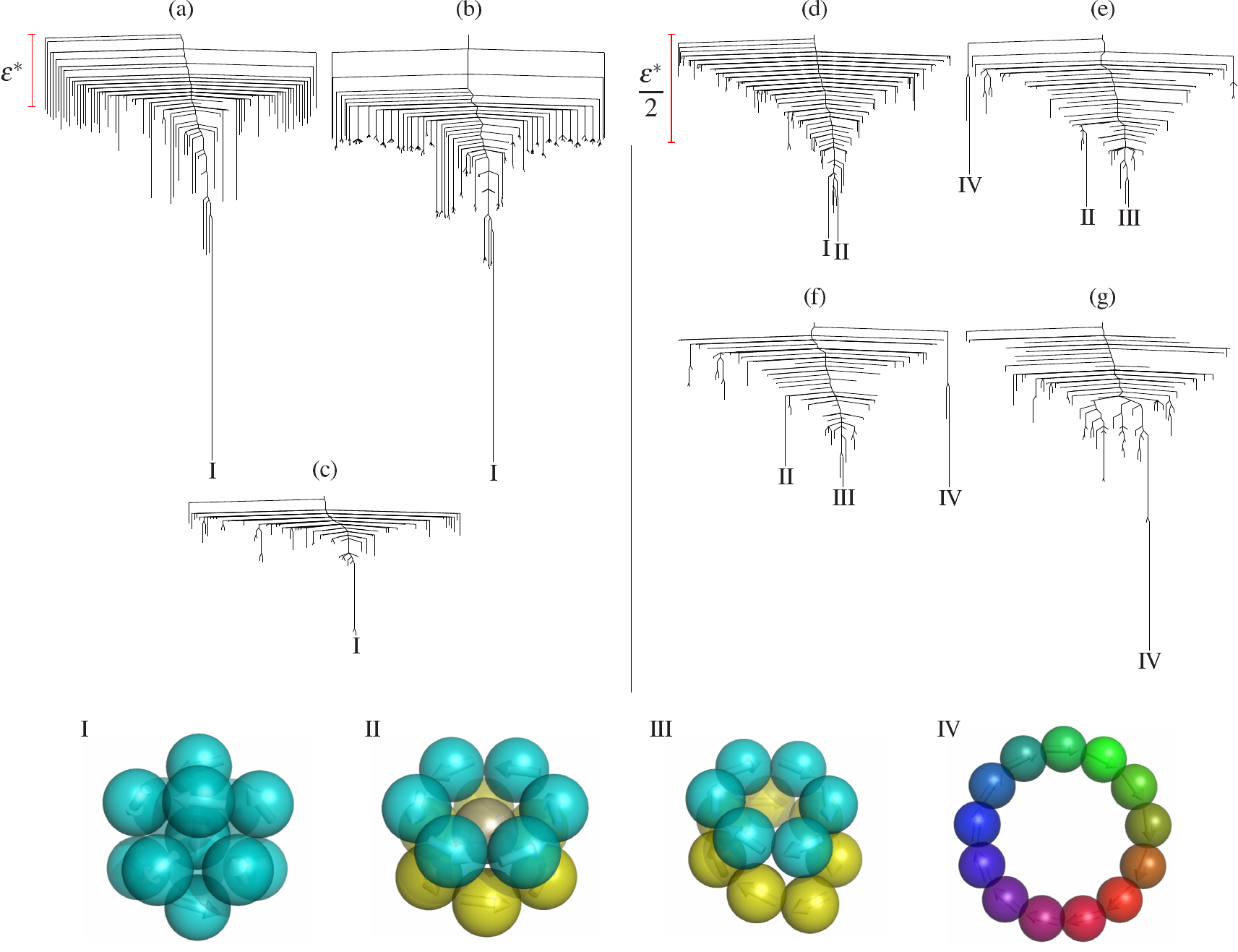}}
\vfill
\caption{
   Upper panel:
      disconnectivity graphs for St$_{13}$ at $\mu=$
      (a) 0;
      (b) 0.2;
      (c) 1;
      (d) 1.42;
      (e) 2.54;
      (f) 2.66;
      (g) 3.6.
      Branches to the 100 lowest energy minima are shown [except in (b) where 300 are shown
for proper comparison with (a)].
   Lower panel:
      global minima for St$_{13}$.
      From left to right, the distorted icosahedron, the hexagonal antiprism, stacked rings, and the single ring.
      Particles are drawn as translucent spheres with arrows pointing along the dipole vector.
      Different chains are distinguished by their colour. Single chains are depicted with smoothly changing colour.
}
\label{fig:T13}
\end{figure*}
as well as at $\mu=0$, 0.2, 1, and 3.6 to investigate the evolution of the potential energy
landscape as a function of dipole strength.

The corresponding disconnectivity graphs are presented in Fig.~\ref{fig:T13}.
For $\mu=0$, the expected single-funnel `palm tree' motif\cite{doyemw99b} is reproduced (Fig.~\ref{fig:T13}a).
This nomenclature refers to the fact that the branches of the graph are vertically rather short on the energy
scale of the pairwise potential and that they are almost all directly connected to the central stem representing the
basin of the global minimum.  These features are indicative of multiple sequences of minima separated by relatively low
barriers, converging on the lowest-energy structure.

Introduction of a weak dipole causes a rapid increase in the number of minima on the landscape, owing to the multiplicity of stable
arrangements of dipoles for each LJ morphology that are easily interconvertible via low barriers (Fig.~\ref{fig:T13}b.)
Aside from the bifurcation of the branches that these arrangements cause, the form of the graph remains the same.
When $\mu$ is increased to 1, elements of this splitting are still present (Fig.~\ref{fig:T13}c).
Two stable structures with practically identical geometries and energies, but different arrangements of dipoles, lie substantially
lower in energy than any other structure.
Morphologies that do not exist on the LJ landscape have begun to emerge and, in terms of the optimal pair energy $\epsilon^*$,
the funnel appears to be shallower.

\begin{figure}[tb!]
\centerline{\includegraphics[width=0.45\textwidth]{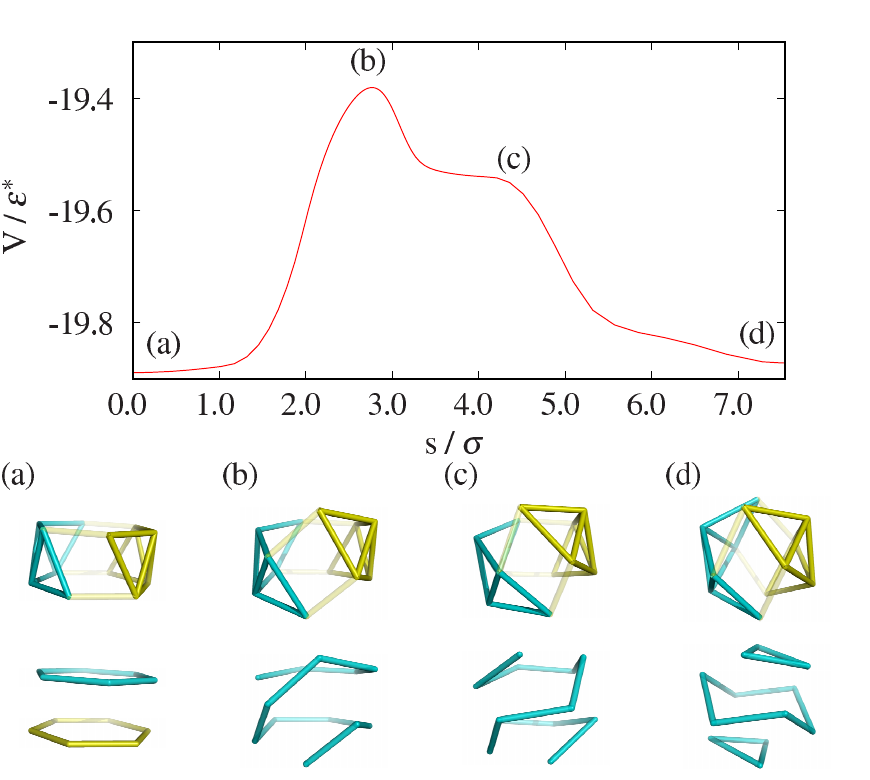}}
\vfill
\caption[EofS13A]{
   Upper panel:
      A path between the hexagonal antiprism and the distorted icosahedron for the St$_{13}$ cluster at $\mu=1.42$.
      The graph shows the energy divided by the pair energy, $V/\epsilon^*$, as a function of the integrated path length, $s$.
   Lower panel:
      Structures at labelled points on the path:
      top, in terms of the dipolar particles discussed in the text,
      and bottom, the dipole network.  (The central particle, which does not participate in dipole bonding, is not shown.)
}
\label{fig:EofS13A}
\end{figure}

At $\mu=1.42$, the global minimum changes from the distorted icosahedron to the hexagonal antiprism (Fig.~\ref{fig:T13}d.)
The disconnectivity graph is still, broadly speaking, a palm tree but now has two low-lying minima that are separated by
a substantial barrier.
The lowest-energy single-step rearrangement in the database
that connects the corresponding minima is presented in Fig.~\ref{fig:EofS13A}.
The coordinate $s$ in this plot is the integrated distance along the pathway between the initial and final structures
in the $3N$-dimensional Euclidean space of the particle positions.  The two portions of a pathway from a transition
state to its directly connected minima are defined by the approximately steepest-descent minimisations initiated
from the transition state along the unstable mode.  The orientations of the dipole are not included
in $s$ and pure rotations of dipoles therefore do not contribute to the apparent length of the path.
The interconversion mechanism is the well-known diamond--square--diamond rearrangement (DSD), proposed by Lipscomb,\cite{DSD}
which describes rearrangements of boranes,\cite{DSDboranes} carboranes,\cite{DSDcarboranes} and
metallaboranes.\cite{DSDmetallaboranes1,DSDmetallaboranes2}
\begin{figure}[tb!]
\centerline{\includegraphics[width=0.45\textwidth]{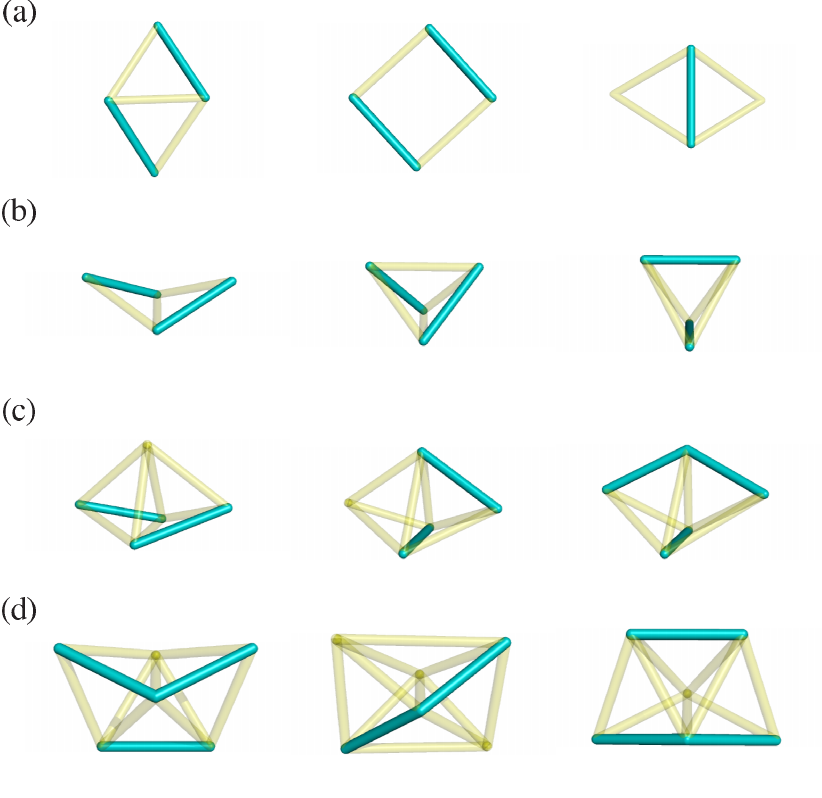}}
\vfill
\caption{
   Rearrangement mechanisms for St$_{13}$ clusters as minimum-transition state-minimum triples:
   (a) the diamond-square-diamond (DSD) rearrangement;
   (b) the butterfly-tetrahedron (BT$_{d}$) rearrangement;
   (c) the closed butterfly-double tetrahdron (B$_{c}$T$_{d}^{2}$) rearrangement;
   (d) the double BT$_{d}$ (BT$_{d}$T$_{d}$B) rearrangement.
   Transparent tubes show the underlying geometry, and filled tubes show dipole bonds.
   \label{fig:mparticle}
}
\end{figure}
It is also an important mechanism for LJ clusters.\cite{uppenbrinkw91}
The pathway contains two consecutive DSD rearrangements, which successively cleave both rings of the hexagonal antiprism at
diametrically opposite points.
To visualise this mechanism it is helpful to think of the hexagonal antiprism as a ring of twelve
edge-sharing triangles.  The DSD rearrangement rotates two opposite segments, each of four triangles,
with respect to one another about a line connecting their centres.  In the final
structure (Fig.~\ref{fig:EofS13A}d) the two segments now clasp each other in a relative orientation
that is orthogonal to where they started (Fig.~\ref{fig:EofS13A}a).

Dipolar `bonding' adds a new dimension to the DSD rearrangement, illustrated in Fig.~\ref{fig:mparticle}a.
Pairs of dipoles begin aligned along opposite edges of the diamond face.
As the long diagonal contracts and the short diagonal elongates, the approaching dipoles reorient so that in the product they are
aligned along the new short diagonal.
The effect is to `break' the bonds along the edges and create a bond along the diagonal of the diamond.
In the rearrangement between the distorted icosahedron and hexagonal antiprism of St$_{13}$, this process breaks one bond
in each ring and forms an interchain contact between two of the free
ends.  Along the pathway there exists a single twelve-particle helix (Fig.~\ref{fig:EofS13A}c)
wrapped around the central particle on the downhill path to the distorted icosahedron, which has
three closed loops of dipoles.

At $\mu=2.54$, the global minimum changes again, from the hexagonal antiprism to two stacked rings (Fig.~\ref{fig:T13}e).
Compared to the first change of global minimum at $\mu=1.42$, the landscape at this second change has more double-funnel
character, since each of the two low-lying minima has a number of other minima associated with its main branch.
In addition, the thirteen-membered planar ring, which is the global minimum in the strong dipole limit, now appears as a metastable
structure connected to the main stem of the graph at high energy.  The associated region of the energy landscape may be
considered as a third funnel.

The pathway connecting the lowest two minima is more complex than at the previous change,
containing five transition states, and traversing a variety of links and coils.
Starting from the stacked rings (structure III in Fig.~\ref{fig:T13}), the first three steps
convert the unlink to a coil, the coil to a link, and finally the link to a different coil.
Each of these transformations occurs by the butterfly--tetrahedron rearrangement (BT$_{d}$),
proposed by Wales \emph{et al.}\cite{walesml89}
It is somewhat similar to Johnson's edge-bridging mechanism\cite{Johnson86} in terms of the
centres of mass: a butterfly moiety closes to form a tetrahedron.
However, in the context of dipolar particles, this underlying structural change allows
recombination of chains, interconverting topological isomers.
This process is illustrated in Fig.~\ref{fig:mparticle}b.
Similarly to the DSD rearrangement, dipoles begin aligned along opposite edges of the butterfly.
In contrast to the DSD process, the long diagonal contracts, but the short diagonal does not
elongate significantly.  As the incoming particles approach, the dipoles reorient to be along
the interparticle vector, and a bond forms along the nascent tetrahedral edge.
Simultaneously the dipoles at the hinge vertices align along the opposite edge.
In this fashion, parallel chain segments are converted into perpendicular ones, creating
a crossing in the chain of dipoles while keeping the number of dipole bonds conserved.

The remaining steps of the interconversion from stacked rings to hexagonal antiprism
involve rearrangements related to BT$_{d}$.  First, a coil is converted to a structure
containing a five-membered ring and a seven-membered ring with a particle in the centre (5,1,7).
This process is difficult to describe in purely geometrical terms, but insight may be
gained by considering which bonds are formed or broken in the process.
This analysis allows us to describe what we will call the closed butterfly--double tetrahedron
rearrangement (B$_{c}$T$_{d}^{2}$), illustrated in Fig.~\ref{fig:mparticle}c.
We begin with a face-capped or `closed' butterfly arrangement of particles, where bonds
are formed along the same edges as are found in the BT$_{d}$ process.
The butterfly closes around this particle, one wing at a time, forming bonds between the
wing-tips and the capping particle, and along the hinge.
The result can be thought of as two face-sharing tetrahedra.
Comparison of Fig.~\ref{fig:mparticle}c and \ref{fig:mparticle}b reveals the similarity
in the overall changes to the dipole network.  In this way a particle is released from
two face-sharing tetrahedra in the coil, which becomes the central particle in the 5,1,7
ring system.

The final step, in which the 5,1,7 system is converted to the hexagonal antiprism,
is a concerted double BT$_{d}$, or BT$_{d}$T$_{d}$B, rearrangement, illustrated in
Fig.~\ref{fig:mparticle}d.  In terms of the particle centres, a butterfly and a
tetrahedron sharing the hinge edge simultaneously close and open, interconverting.
It differs from the other examples in that the butterfly's dipole bonds are not
opposite each other, but adjacent, and the tetrahedra have bonds only along
the edge that is broken in the butterfly.  In the context of the St$_{13}$ structure,
the rearrangement of this fragment effects a particle exchange between the rings.
At the transition state of the whole structure, none of the tetrahedral bonds and
one butterfly bond for each of the rings exist, forming a twelve-particle helix.

The global minimum in the high-dipole limit is a single thirteen-particle ring.  This structure
first replaces the stacked rings as the Stockmayer global minimum at $\mu=2.66$.
The disconnectivity graph (Fig.~\ref{fig:T13}f.) shows distinct funnels
for the two competing structures, but the
stacked ring funnel is associated with a larger region of configuration space.
As $\mu$ increases, and the ring funnel becomes increasingly favourable,
the region of configuration space associated with the stacked rings is expected to serve
as a kinetic trap to structural relaxation.

\begin{figure}[tb!]
\centerline{\includegraphics[width=0.45\textwidth]{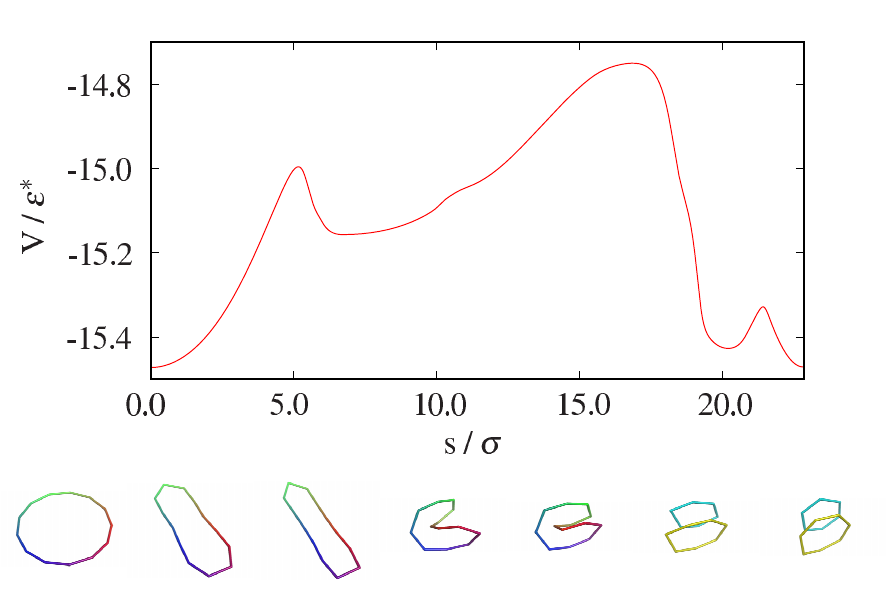}}
\vfill
\caption[EofS13B]{
   Upper panel:
      a path between the ring and stacked rings minima of the St$_{13}$ cluster at $\mu=2.66$.
   Lower panel:
      structures of minima and transition states along the path.
}
\label{fig:EofS13B}
\end{figure}

\begin{figure}[tb!]
\centerline{\includegraphics[width=0.45\textwidth]{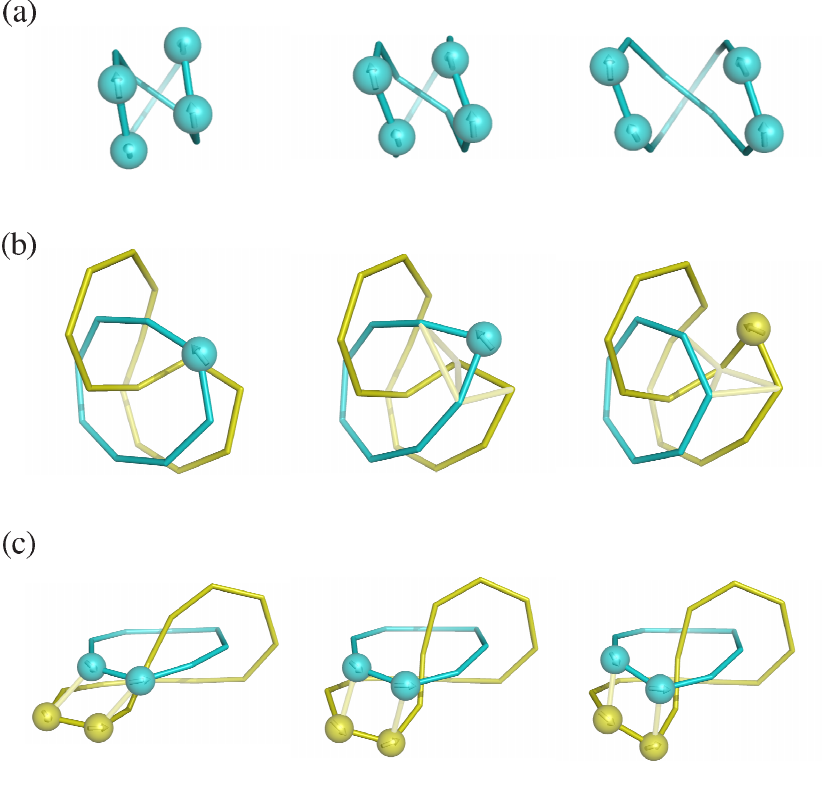}}
\vfill
\caption{
   Chain-based mechanisms:
   (a) a diamond--square--diamond rearrangement;
   (b) a budding rearrangement;
   (c) a DSD rearrangement that facilitates chains moving past one another;
   \label{fig:mchain}
}
\end{figure}

A pathway containing three transition states, which connects these competing minima, is
presented in Fig.~\ref{fig:EofS13B}.
The first two steps flatten the ring into an ellipse and twist it about its centre to form a coil.
Although much of the bending energy penalty is removed, and favourable contacts are made between
particles in the adjacent chains, the sharp bends at the ends of the ellipse are sufficient to
increase its energy with respect to the ring.  As the ellipse twists and folds into a coil,
it is locked in place by a mechanism reminiscent of a DSD rearrangement, which is
illustrated in Fig.~\ref{fig:mchain}a.
The final step in the path is the now familiar BT$_{d}$ rearrangement, which interconverts
the coil and the stacked rings, removing a crossing from the chain.  It is interesting to
note that in the previous pathway, at a smaller value of $\mu$, all of the BT$_{d}$ steps
lowered the energy upon forming the more compact tetrahedral site.
In the current pathway, the situation is reversed, reflecting the increasing stability of
planar moieties as the dipolar contribution to the energy increases.

At large $\mu$, the ring is firmly established as the global minimum.
It is separated from the next lowest minimum, the ellipse, by $0.59\epsilon^{*}$, and
from the third lowest energy structure by $0.78\epsilon^{*}$.

\subsection{St$_{21}$ clusters}

\begin{figure}[!tb]
\centerline{\includegraphics[width=0.45\textwidth]{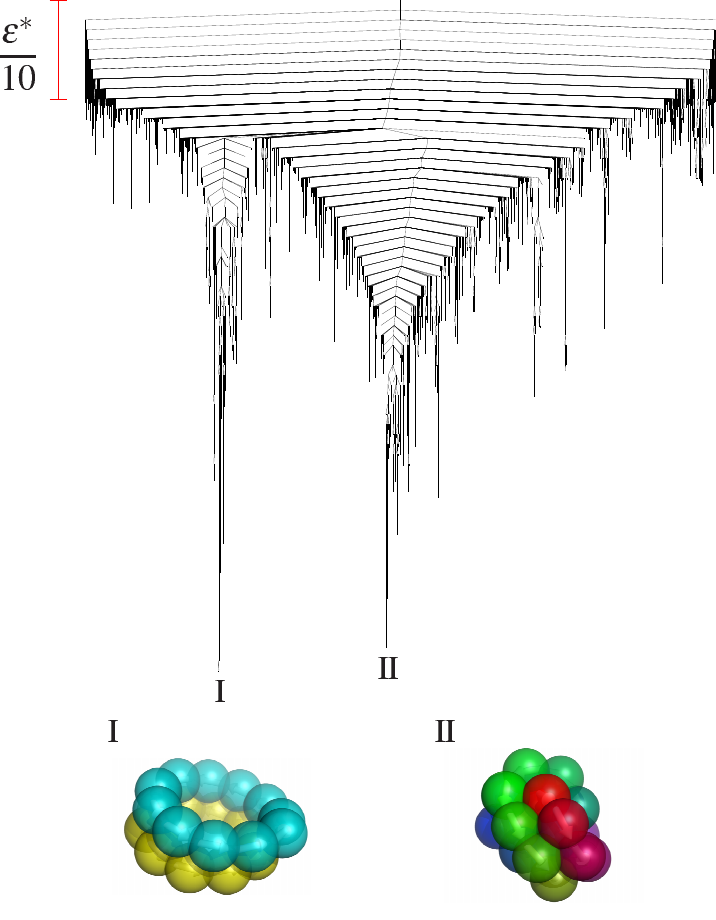}}
\vfill
\caption{
   Upper panel:
      disconnectivity graph for St$_{21}$ at $\mu=$2.90. Branches to the 1000 lowest energy minima are shown.
   Lower panel:
      structures of the stacked ring (I) and trefoil knot (II) minima.
}
\label{fig:T21}
\end{figure}

The most remarkable structures that appear as global minima for Stockmayer clusters
are knotted geometries.\cite{MillerW05}
The smallest cluster where a knot becomes energetically optimal is
St$_{21}$.  The global minimum of this cluster is topologically a trefoil knot
over the range $1.7\leq\mu<2.9$, beyond which a pair of stacked twelve and
thirteen particle rings becomes more favourable.
The disconnectivity graph at $\mu=2.9$ is presented in Fig.~\ref{fig:T21},
and exhibits a pronounced double funnel.
As with the graph at $\mu=2.66$ for St$_{13}$, the wider funnel is associated with the
more compact structure---in this case, the knot.

\begin{figure}[tb!]
\centerline{\includegraphics[width=0.45\textwidth]{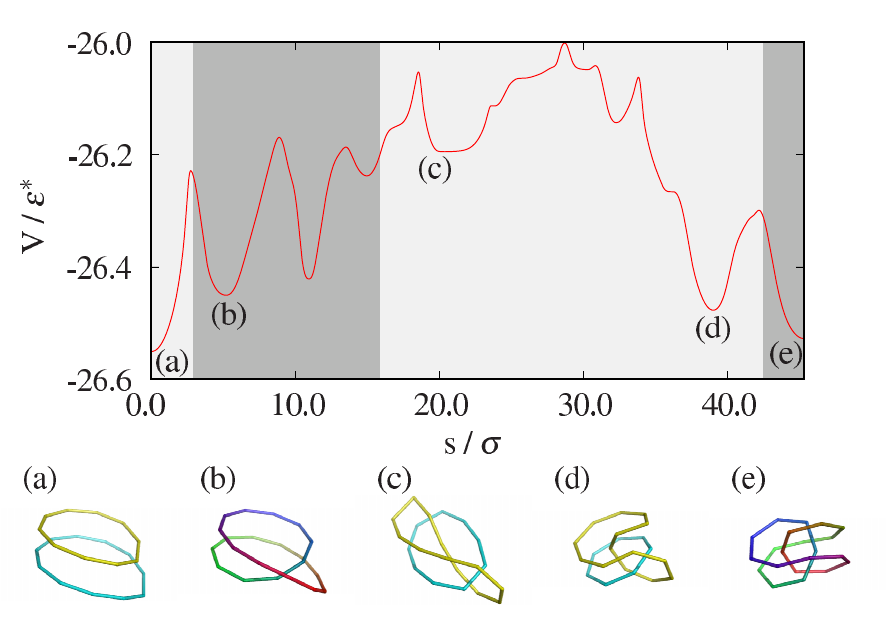}}
\vfill
\caption[EofS21]{
   Upper panel:
      a pathway between the stacked rings and trefoil knot minima of St$_{21}$.
      The figure is shaded according to the number of chains in the structure (light=1, dark=2.)
   Lower panel:
      selected structures from the path:
      (a) the stacked rings;
      (b) a coil;
      (c) a link (9 and 12 particles);
      (d) a link (7 and 14 particles);
      (e) the trefoil knot.
}
\label{fig:EofS21}
\end{figure}

A low-energy pathway connecting these structures is presented in Fig.~\ref{fig:EofS21}.
The pathway contains eleven transition states, and mostly involves rearrangement of, and
particle exchange between, rings in a link.  The first such conversion, from an unlink with
ten and eleven particle rings to a link with nine and twelve particle rings, is
achieved in three transition states.  First, the unlink rearranges to a coil by the BT$_{d}$
mechanism.  This coil contracts to a more compact bundle in a similar fashion to the St$_{13}$ ring.
The coil itself then twists, facilitated by coupled inter-chain DSD rearrangements, which
enable chains to move past one another.
Finally, the coil converts to the nine- and twelve-particle link by a second BT$_{d}$
rearrangement, which conserves a twist in the larger ring (Fig.~\ref{fig:EofS21}c).

Two further exchanges, to form links of eight- and thirteen-particle rings and then the
link of seven- and fourteen-particle rings, occur by a new mechanism,
whereby a particle is smoothly transferred between adjacent chains.
This `budding' rearrangement is illustrated in Fig.~\ref{fig:mchain}b.
A three-particle section of a chain bends sharply, expelling a particle and forming a new
connection between the free ends.  The released particle is incorporated into an adjacent
chain, which opens as the first one closes.  At the transition state, the five sites
are arranged in a closed butterfly geometry.
The process can be thought of as an edge-bridging particle passing to the opposite edge
via a face-bridged transition state.  The latter exchanges are separated by coiling of
the larger ring around the smaller by means of DSD rearrangements, shown in
Fig.~\ref{fig:mchain}c, and the rings are disposed such that a final BT$_{d}$
rearrangement produces the trefoil knot.

\subsection{St$_{38}$ clusters}
The LJ$_{38}$ cluster is unusual in that the global minimum is a truncated octahedron,
and not based on an icosahedron.\cite{Pillardy95a}
The landscape is double-funneled,\cite{DoyeMW99a} and the lowest energy icosahedral minimum,
while significantly higher in energy than the global
minimum, is associated with a much larger region of configuration space.
As such, the system provides a useful test of global optimisation procedures.

St$_{38}$ exhibits a more complex knotted global minimum than the St$_{21}$ trefoil
in the range $1.6<\mu<2.2$.\cite{MillerW05}  This St$_{38}$ knot has eight crossings
in its minimal projection and has the topology $8_{19}$ in Rolfsen's
notation.\cite{Rolfsen76}

\begin{figure}[!tb]
\centerline{\includegraphics[width=0.45\textwidth]{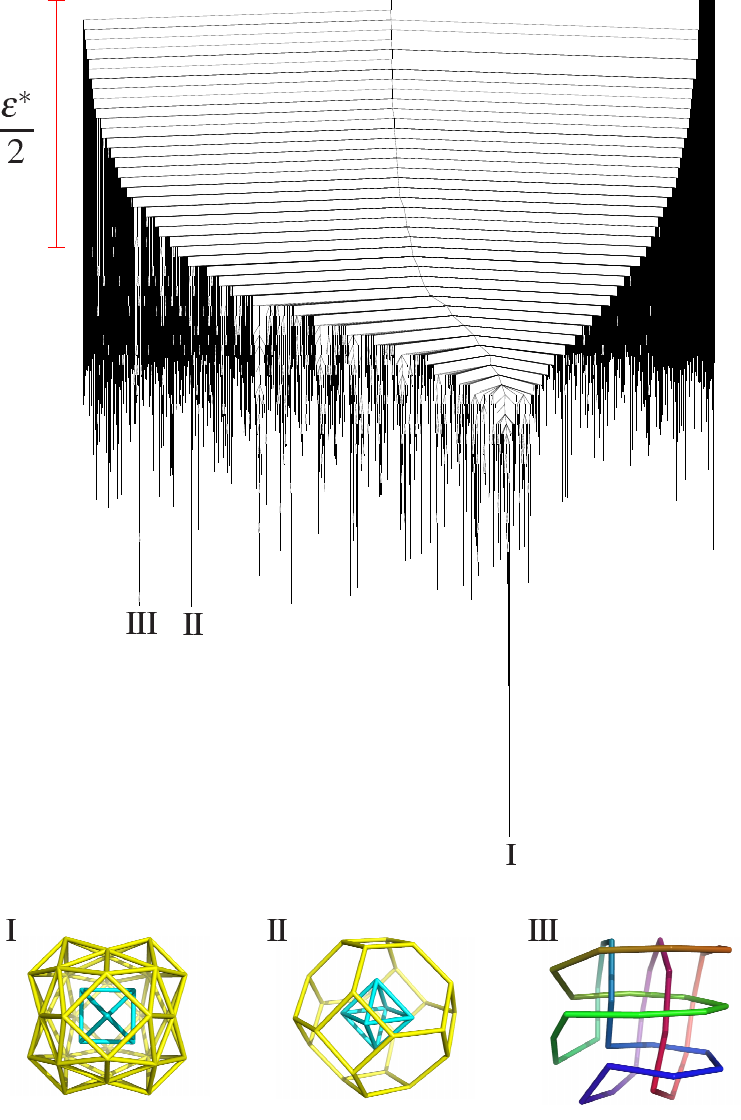}}
\vfill
\caption{
   Upper panel:
      disconnectivity graph for St$_{38}$ at $\mu=1.22$. Branches to the 1000 lowest energy minima are shown.
   Lower panel:
      the $D_{4h}$ global minimum (I) and the $O_{h}$ fcc minimum (II), showing the core-shell structure, and the $8_{19}$ knot
      minimum, showing the dipole chain.
}
\label{fig:T38}
\end{figure}

The Lennard-Jones truncated octahedron ceases to be the global minimum at $\mu=0.8$.
For values of $\mu$ between 0.8 and 1.6, the truncated octahedron and knot---both
metastable---are close in energy.
Databases of minima and transition states have been constructed at $\mu=1.22$, where
the two morphologies are closest in energy.  The disconnectivity graph is presented
in Fig.~\ref{fig:T38}.
It has the form of a gently sloping single funnel, with a form somewhere between a
`weeping willow' and a palm tree.\cite{WalesMW98}
(The willow tree has longer vertical branches connected to the main stem,
indicating multiple descending pathways towards the global minimum, like the
palm tree, but with higher barriers between adjacent minima.)
The global minimum has $D_{4h}$ symmetry and is
depicted in Fig.~\ref{fig:T38}.

\begin{figure}[tb!]
\centerline{\includegraphics[width=0.45\textwidth]{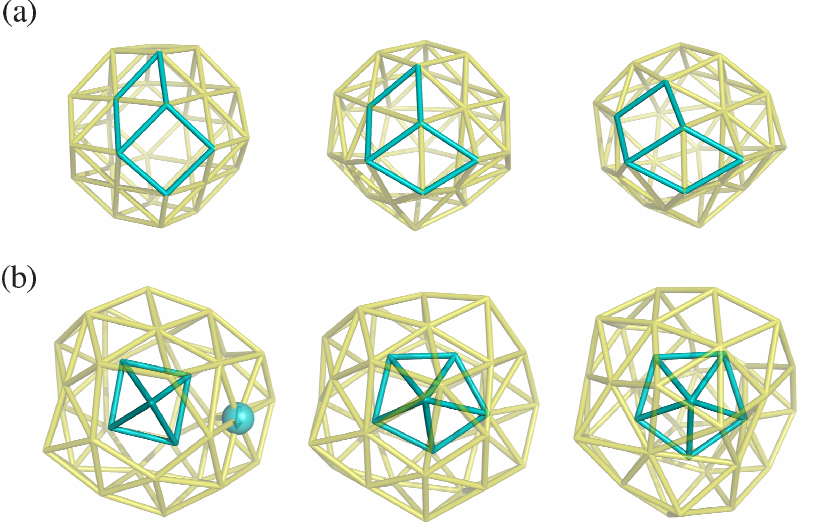}}
\vfill
\caption{
Steps in the rearrangement of St$_{38}$ from a truncated octahedron
(structure II in Fig.~\ref{fig:T38}) to an $8_{19}$ knot (structure III).
(a) A square-diamond--diamond-square (SDDS) rearrangement on the surface
of the truncated octahedron (structure II in Fig.~\ref{fig:T38}).
(b) A core-shell rearrangement in the final step of the pathway.
Filled tubes highlight the geometry of the rearranging sites.
\label{fig:m38}
}
\end{figure}

The truncated octahedron and knot structures can be interconverted by a pathway
containing just three transition states.
Multiple concerted surface rearrangements occur in the first two steps, an example of
which is presented in Fig.~\ref{fig:m38}a.
The square faces of the truncated octahedron minimum permit the previously
characterised\cite{walesm89b} square-diamond--diamond-square (SDDS) mechanism to occur.
The final step is an intershell transfer of a particle from the surface to the core,
expanding the octahedron to a pentagonal bipyramid (Fig.~\ref{fig:m38}b).
Since we are considering a relatively weak dipole, the absence of dipole-dominated
mechanisms, such as BT$_{d}$, is not surprising.  The 8$_{19}$ knot is still compact,
but dipole bonding stabilises the open faces required to arrange the particles in
such a way that the dipoles may orient to form a knot.

\section{Conclusions}

Clusters of spherical particles bound by simple isotropic potentials like Lennard-Jones
and Morse are generally dominated by compact---often icosahedral---structures.
The anisotropic Stockmayer potential differs from such cases
because of the particles' tendency
to form chains.  The fact that chain-like motifs can readily be identified
introduces the new consideration of topology when characterising
structure and rearrangements.  Since the connectivity of particles
within the chains is not fixed (as it is in a polymer),
rearrangements may alter the topology.

Some rearrangement mechanisms in Stockmayer clusters resemble those found in
clusters bound by other potentials.  For example, the diamond--square--diamond
rearrangement, which is common in clusters bound by isotropic potentials, also
occurs in Stockmayer clusters.  However, this familiar mechanism gains additional significance
because of the change in orientation of the dipole vectors, which can allow chains to be
disconnected in one direction and to become connected in another.  Similarly,
edge-bridging mechanisms seen in other clusters can change the
topology of Stockmayer clusters.

In addition to mechanisms that change the connectivity between chains while
preserving the association of each particle with its chain, there are
budding processes.  Here, a particle is smoothly ejected from one chain and
absorbed into another, thereby allowing chains to change in size.

These mechanisms vary in importance with the strength of the dipole moment.  A crucial
consideration is whether or not the pathway conserves the number of head-to-tail
connections between dipoles.  The diamond--square--diamond mechanism exchanges two parallel
connections for one in an orthogonal orientation, and so is more favourable at lower
dipole strengths, where the loss of a dipole--dipole bond does not incur a large
energetic penalty.  The butterfly--tetrahedron mechanism and some of its variants
preserve the number of bonds while interconverting topological isomers, and
become more important at higher dipole strengths.

Stockmayer clusters are frustrated systems in which the isotropic Lennard-Jones
part of the potential drives the structure towards compact highly-coordinated
arrangements, while the dipolar interactions favour chain-like motifs.
The analysis of structures and rearrangement mechanisms that we have presented here
for St$_{13}$, St$_{21}$, and St$_{38}$ illustrates a variety of topology-changing
events that offer a compromise between the competing terms in the potential.

Experimentally, clusters of colloids with a dipolar component to their interactions
are being studied with increasing levels of control, and are being characterised in
greater detail.  While chain- and loop-based structures currently predominate, it
should be possible to explore more compact clusters as well.  The knotted topologies
and other interesting structures that arise for Stockmayer particles are obtained
by adjusting the relative strengths of the dipolar and isotropic parts of the potential.
This flexibility can be achieved in suspensions of dipolar colloids by
introducing tunable, depletion-induced, isotropic attraction via an additional component in
the suspension.\cite{Ilett95a}  The isotropic attraction can also be controlled through
the surface chemistry of the colloids.\cite{Keng07a}
In both cases, the repulsive core of the particles is likely to be considerably harder
than the $r^{-12}$ term in Eq.~(\ref{eq:stockmayer}).  However, it is the
presence of an isotropic attractive tail and its competition with the dipolar
interactions that are crucial in generating a rich selection
of topologies.

A number of open questions remain in the context of clusters and molecules with
non-trivial topology.  These issues include the maximum complexity (number of crossings)
that can be attained in a given system, and the precise factors determining
the favoured topology where a choice exists.  The influence of the interparticle
potential and any fixed connectivity between particles will be important in this regard, and
these considerations make contact with other knot-forming systems, such as proteins and 
synthetic organic molecules.  The study of idealised model systems, such as the work presented 
here, should be a fruitful way of of gaining insight into the general phenomenon of knot formation.

\section*{Acknowledgments}
The authors are grateful to EPSRC and to the European Research Council for financial
support.  D.C.~acknowledges support through
an Oppenheimer Research Fellowship from the University of Cambridge.

\footnotesize{
\providecommand*{\mcitethebibliography}{\thebibliography}
\csname @ifundefined\endcsname{endmcitethebibliography}
{\let\endmcitethebibliography\endthebibliography}{}

}

\end{document}